\documentclass[default]{sn-jnl}
\usepackage{siunitx}
\usepackage[version=4]{mhchem}
\usepackage[capitalise,noabbrev]{cleveref}
\usepackage[superscript]{cite}

\begin{document}

\title[]{Accelerating point defect simulations using data-driven and machine learning approaches}

\author[1]{\fnm{Arun} \sur{Mannodi-Kanakkithodi}}\email{amannodi@purdue.edu}

\author[2]{\fnm{Menglin} \sur{Huang}}\email{menglinhuang@fudan.edu.cn}

\author[3]{\fnm{Prashun} \sur{Gorai}}\email{goraip@rpi.edu}

\author[4]{\fnm{Seán} \sur{R. Kavanagh}}\email{sk2045@cam.ac.uk}

\affil[1]{School of Materials Engineering, Purdue University, West Lafayette, IN 47907, USA}

\affil[2]{Key Laboratory of Computational Physical Sciences (MOE), Fudan University, Shanghai 200433, China}

\affil[3]{Department of Chemical \& Biological Engineering, Rensselaer Polytechnic Institute, Troy, NY 12180, USA}

\affil[4]{Yusuf Hamied Department of Chemistry, University of Cambridge, Lensfield Rd, Cambridge CB2 1EW, United Kingdom}

\abstract{Point defects in solid-state materials are now routinely simulated using large supercell structures, requiring efficient quantum mechanical solutions. Data-driven and machine learning (ML) models trained on computational data can enable rapid defect property predictions and high-throughput screening. In this article, we provide an overview of prominent efforts to accelerate defect simulations using these approaches. We begin by discussing the motivations for data-driven techniques in defect modeling, and describe efforts over the past decade to use descriptor-based models for rapid screening of defect properties -- most notably in oxides. In particular, we discuss case studies where surrogate models and interatomic potentials were trained on density functional theory (DFT) data, leading to predictions with quantum-mechanical accuracies at a fraction of the cost. In addition to geometry relaxation and formation energy predictions, these interatomic potentials are capable of predicting phonon modes and vibrational free energies to yield defect energetics at finite temperatures -- representing a key frontier for computational defect research. 
We finish with a discussion on how to connect these approaches and their outputs with experimental data, and provide an outlook on this burgeoning sub-field.}


\maketitle
\markboth{}{}

\section*{Introduction}

Characterizing the behavior of point defects in solid-state materials is vital to understanding their properties and limitations. Defects impact many crucial properties of functional materials, such as electronic and ionic conductivity, catalytic activity, electron-hole recombination, optical emission, crack formation, and more, dictating performance for photovoltaics, catalysis, power electronics, energy storage, and quantum technologies \cite{Def-1,Def-2,Def-3}. First principles computations are frequently used to quantitatively predict which native and extrinsic point defects may form in solids, and how they are influenced by the temperature, the Fermi level, or chemical growth conditions \cite{Def-4,Def-5,Def-6}. Periodic Density Functional Theory (DFT) is by far the most common electronic structure method employed, given its excellent cost-accuracy trade-off, but other theories such as $GW$ and Quantum Monte Carlo (via embedding) have also been trialed in select cases \cite{kleiner_quantum_2025,chen_accuracy_2017,dernek_realspace_2022}.
These simulations reveal the equilibrium polarity and conductivity in semiconductors or insulators ($p$-type, $n$-type, or intrinsic), based on the formation energies of charged native vacancies, self-interstitials, and anti-site substitutions, extrinsic substitutional or interstitial impurities or dopants, and defect complexes that involve two or more of these individual defects in proximity \cite{Dop-1,Dop-2,Dop-3}. \\

High-throughput DFT has been used to generate datasets of defect formation energies, charge transition levels, and related properties, with the goal of screening and designing suitable functional dopants and/or defect-tolerant materials \cite{ML-Wan,ML-Deml-1,Mannodi-EES}. However, DFT computations for point defects involve large supercells, image charge corrections, and advanced functionals (or corrections) that yield suitable band edges and gaps, making them expensive and prohibitive in many cases. A variety of data-driven and machine learning (ML) approaches have been proposed for accelerating defect simulations and overcoming this bottleneck. This includes descriptor-based models where defect properties are predicted using an input representing the compound, the defect site, and the defect species, as well as methods involving machine-learned force fields (MLFFs), which attempt to directly reproduce energies and atomic forces from quantum-mechanical simulations. \\

``DFT-ML" is now ubiquitous; most high-throughput computations involve some acceleration with ML, with final prediction and screening retaining DFT accuracy (at a desired level of theory determined by the training dataset) but at a much lower computational expense. Defect simulations have been no strangers to this trend. Data-driven and machine-learning models have been trained to obtain important bulk or local descriptors, or surrogate quantities, that correlate with defect properties; to directly predict formation energies and defect levels; to optimize the geometries of charged and neutral defect configurations; and to design low energy and electronically active defects and dopants across a variety of crystalline materials. The advent of ML-augmented defect simulations is potentially a game-changer for studying and designing materials with tailored defect and doping properties. \\ 

Embodying the spirit of Frenkel and paying tribute to the many giants who established new frontiers in point defect simulations, we present a perspective on data-driven and machine learning approaches for understanding the properties of defects and dopants in solid-state materials. This article will emphasize how these methods are beginning to be adopted for high-throughput screening; to extend materials design to predictions of defect tolerance and dopability; to enable prediction of defect-derived properties such as ion migration pathways and energies, defect phonon modes, carrier capture rates, and carrier concentrations; and to interpret experimentally measured peaks and energy levels. In the next few sections, we discuss DFT simulations for point defects and the need for data-driven screening and ML acceleration, and present a selection of case studies where these concepts were brought to life. We discuss the strengths and limitations of current approaches in predicting defect properties and how they connect with experiments. We end with an outlook on what the future of ML-guided defect simulations looks like, the unique advantages offered by ML-driven defect predictions, and where we can expect to be in 5-10 years and beyond. \\

\section*{Simulating point defects using DFT: Theoretical background and the need for machine learning}

Point defects are now routinely simulated using quantum-mechanical methods \cite{freysoldt_firstprinciples_2014}, of which DFT is by far the most popular. While a number of simulation procedures exist, such as infinite-crystal embedding and cluster-based approaches, the supercell approach has emerged as the dominant method of choice -- benefiting from the speed and reliability of periodic DFT codes.
Here, defects are placed in large periodic supercells to simulate isolated, dilute defects, with corrections often applied to account for spurious finite-size interactions. \\

Among the key properties of interest in defect calculations are the atomic geometry of the defect, its formation energy and its electronic energy levels, which dictate their impact on relevant macroscopic properties. The defect formation energy (E$^f$) and defect charge transition levels ($\epsilon$(q$_1$/q$_2$)) are calculated using \textbf{Eq. \ref{Eq:DFE}} and \textbf{Eq. \ref{Eq:CTL}}, respectively presented below:

\begin{equation}
E^f(D^q, E_F) = E(D^q) - E(\text{bulk}) + \sum_i n_i \mu_i + q(E_F + E_{\text{VBM}}) + E_{\text{corr}}
\label{Eq:DFE}
\end{equation}

\begin{equation}
\varepsilon\left(q_1 / q_2\right)=\frac{E^f\left(q_1\right)-E^f\left(q_2\right)}{q_2-q_1}
\label{Eq:CTL}
\end{equation} \\

In these equations, E(D$^q$) is the total energy of the supercell containing the defect in a charge state \(q\), \(E(\text{bulk})\) is the total energy of the pristine bulk supercell, \(n_i\) is the number of atoms of species \(i\) added (negative) or removed (positive) to form the defect, \(\mu_i\) is the chemical potential of species \(i\) participating in the defect referenced to its elemental standard state, q$_1$ and q$_2$ represent the defect charges, \(E_F\) is the Fermi level which represents the electronic chemical potential and is typically given relative to the valence band maximum (VBM), $E_{\text{VBM}}$ is the electronic eigenvalue of the VBM, and \(E_{\text{corr}}\) is the correction term for the spurious electrostatic interactions in charged defects. $\epsilon$(q$_{1}$/q$_{2}$) corresponds to the $E_{F}$ value where the defect ground-state transitions from charge state q$_{1}$ to q$_{2}$ (independent of $\mu$ conditions) and is calculated using \textbf{Eq. \ref{Eq:CTL}}. \\

The temperature-dependent equilibrium concentration of a point defect in a charge state $q$ is determined by the formation energy, following the equation (at the dilute limit):

\begin{equation}
C_X = N_X g_X^q \exp\left(\frac{-E_f(X^q)}{k_BT}\right)
\label{Eq:conc}
\end{equation} \\

Here, $N_X$ is the number of available lattice sites for incorporation of defect $X^q$ and $g_X^q$ is the configurational degeneracy of defect $X$ in a charge state $q$; \textcolor{black}{these numbers are normalized appropriately, such as by volume}.  The concentrations of available electrons in the conduction band and holes in the valence band can additionally be calculated using the Fermi-Dirac distribution and the electronic density of states (DOS). Then, to compute the equilibrium Fermi level $E_F^{\textrm{eq}}$, an iterative search is employed between the VBM and the CBM to find the point of charge neutrality, i.e., where the concentrations of electrons and negatively charged (acceptor) defects cancels out the concentrations of holes and positively charged (donor) defects \cite{squires_guidelines_2026}. A highly accurate electronic structure calculation for the host compound is crucial to obtaining the correct band gap and band edges, and its expense is typically far lower than the charged defect supercell calculations. \\

\textcolor{black}{The above equations typically ignore the temperature-dependence of the electronic band edges and chemical potentials, which may lead to inaccurate defect energies and thus have consequences for computed concentrations and predicted defect impacts.\cite{Arnab_2025,kavanagh_accurately_2024,mosquera-lois_imperfections_2023}
Furthermore, finite-size corrections with appropriate dielectric screening are crucial for the accurate prediction of charged defect levels and energies in typical simulation supercell sizes.\cite{freysoldt_firstprinciples_2014,squires_guidelines_2026} Eq. \ref{Eq:conc} is valid for the dilute defect limit, while extensions are necessary for quantitative accuracy with high defect concentrations, where defect-defect interactions and complex association become significant.\cite{Biswas_2009,krasikov_defect_2017,cen_cation_2023,krasikov_thermodynamic_2018}} \\

\begin{figure*}[ht]
\centering
\includegraphics[width=\linewidth]{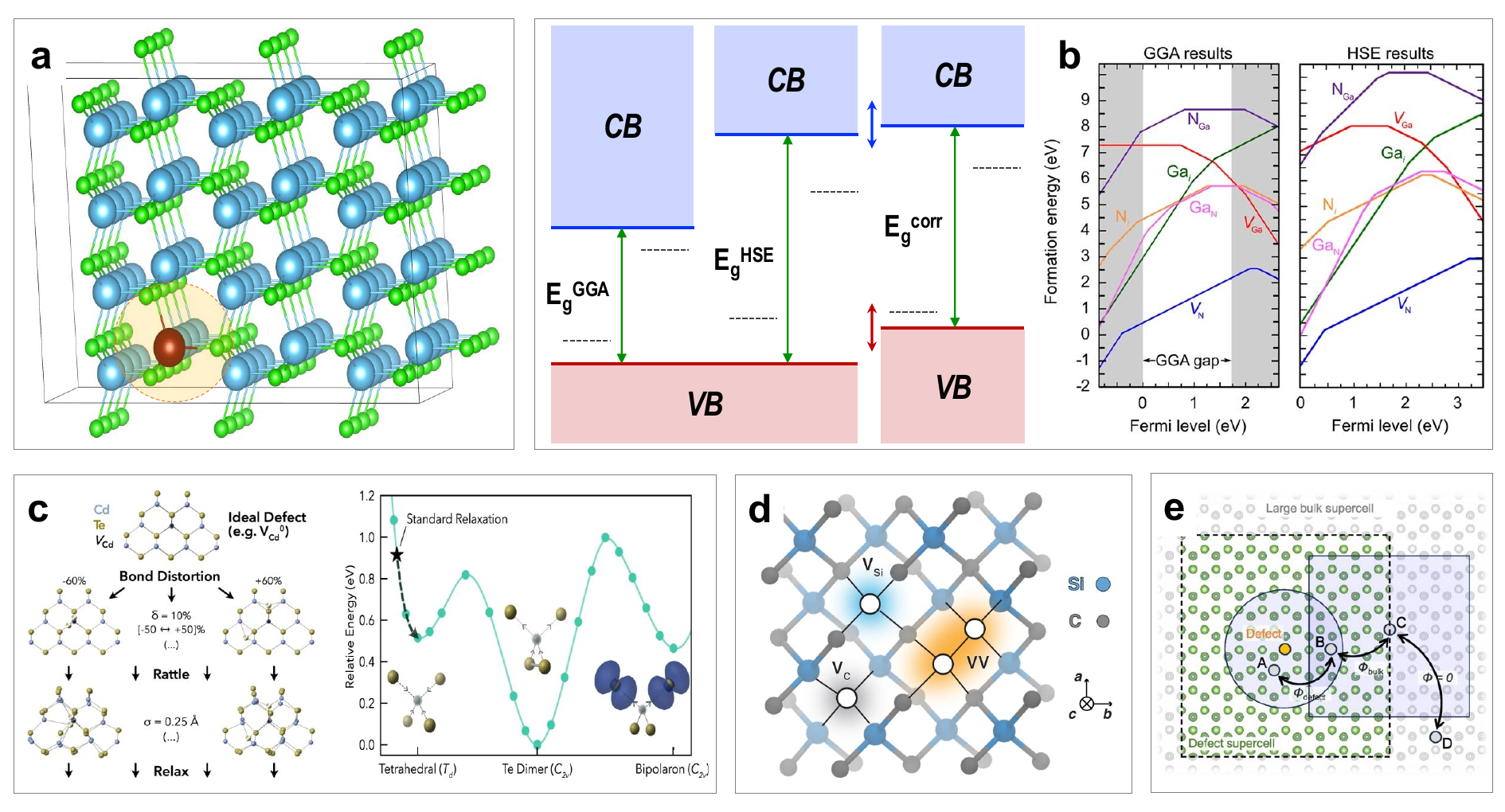}
\caption{\textbf{Important considerations for defect simulations with DFT.} \textbf{(a)} A typical defect-containing supercell, shown as an example for a substitional defect in Wurtzite-phase AlN. \textbf{(b)} The level of theory (e.g., semi-local GGA vs non-local hybrid DFT) determines accuracy of the band gap and band edges, which could also be physically shifted to match experimental values \cite{Mannodi-Patterns,Broberg}. VB = valence band, CB = conduction band, E$_g$ = band gap, and the dashed lines are hypothetical defect levels. Also pictured is a comparison between defect formation energy diagrams in GaN from GGA and HSE, reproduced with permission from Lyons et al. \cite{Lyons-GaN}. \textbf{(c)} Intentional symmetry-breaking in defect coordination environments can identify lower-energy minima; reproduced with permission from Mosquera-Lois et al. \cite{localmin}. \textbf{(d)} Example of a defect complex (a double vacancy) in SiC, reproduced with permission from Lee et al. \cite{Lee-SiC}. \textbf{(e)} Constructing force constants from a defect supercell for calculating vibrational properties; reproduced with permission from Zhou et al. \cite{zhou2025onedefectpotentialstrategy}.}
\label{fig:Intro_Figure}
\end{figure*}

For achieving accurate defect descriptions, there are several important considerations, which are presented in \textbf{Figure \ref{fig:Intro_Figure}} and detailed below: \\

\begin{itemize}

    \item \underline{Large supercell}: Point defects must be simulated in large supercells, ideally extrapolated to the infinite size limit, to account for the dilute nature of defects. Current best practice involves choosing a sufficiently large cell size that best compromises between DFT expense and convergence of the computed defect formation energy. \textcolor{black}{For instance, defect simulations for binary semiconductors which adopt the zincblende phase, such as CdTe, ZnS, and GaAs, frequently employed 64-atom 2$\times$2$\times$2 supercell expansions of the conventional unit cell until maybe a decade ago \cite{Pham}, but 216-atom 3$\times$3$\times$3 supercells or even larger are now more prominent \cite{Flores,Schultz}. Cubic ABX$_3$ perovskites often used 40-atom 2$\times$2$\times$2 supercells, but the recent use of larger supercells have resulted in more accurate defect treatments \cite{Souza}. Computational expense typically scales dramatically with increasing cell size. For instance, DFT cost scales approximately cubically with the number of atoms (electrons), which itself scales cubically with cell dimension: e.g., from a 2$\times$2$\times$2 to a 3$\times$3$\times$3 CdTe supercell, the cubic cell length increases from $\sim$13$\AA$ to $\sim$20$\AA$, while the computation time increases by almost two orders of magnitude.} With modern computing resources that provide efficient computational scaling and ever more accurate ML models, much larger supercell sizes than before are now being used for simulating defects and calculating energy differences. \\
    
    \item \underline{Correct band gap and band edges}: Point defects will often be charged, which means their formation energies vary with the electron chemical potential, making it crucial that the valence and conduction band edges are accurately described. The solution here typically involves using an advanced level of theory, often a hybrid DFT functional with a tailored mixing fraction and spin-orbit coupling if appropriate. Local and semi-local functionals are much faster in computing large supercell defect configurations and thus conducive to high-throughput studies, but come with a number of established failures \cite{squires_guidelines_2026,freysoldt_firstprinciples_2014,Lyons-GaN}. \textbf{Figure \ref{fig:Intro_Figure}b} shows possible band edge shifts to change defect levels and a comparison between defect formation energy diagrams computed for GaN from GGA and HSE \cite{Lyons-GaN}. A clear goal for ML then is to bridge the gap between these levels of theory and/or directly yield electronic and defect predictions at a high level of theory.  \\
    
    \item \underline{Considering many charge states}: Defects in semiconductors and insulators can localize varying amounts of electronic charge at their sites, thus adopting different charge states whose formation energies have different dependence on the Fermi level (\textbf{Eq. \ref{Eq:DFE}}). Neutral defects may exhibit formation energies that are artificially high leading to an incorrect ordering of defects by relative energetics, thus making it necessary to perform charged calculations. It is vital to consider any native defect, dopant, impurity, or defect complex in multiple possible charge states to obtain the complete picture of formation energies and accessible charge states, which further adds to the overall expense. \\
    
    \item \underline{Symmetry-breaking and comprehensive configurational exploration}: It has been shown that without intentional geometry perturbations, defect structure optimization may find itself trapped in local minima and fail to find the true ground state configuration \cite{localmin,arrigoni_evolutionary_2021,lany_anion_2005}. A proposed solution is the creation of several trial configurations with different degrees of perturbations applied around the defect center \cite{kavanagh_doped_2024,mosquera-lois_shakenbreak_2022}, followed by optimization of each to identify stable structures. \textbf{Figure \ref{fig:Intro_Figure}c} shows an example of this approach; applying initial bond distortions and rattling before geometry optimization, which identifies lower-energy dimer and bi-polaron configurations for a Cd vacancy in CdTe \cite{localmin,kavanagh_rapid_2021}. While quantitative accuracy can be reduced to accelerate these relaxations, they can remain a time-intensive task for high-throughput studies, motivating the use of ML force fields for much faster optimization \cite{mosquera-lois_machinelearning_2024,kavanagh_identifying_2025,arrigoni_evolutionary_2021}. \\
    
    \item \underline{Simulating all essential reference phases}: To calculate chemical potentials of all relevant species at different chemical growth extremes (e.g., cation-rich and anion-rich conditions), the energies of all alternative and elemental phases are necessary. \textcolor{black}{Using these energies, multiple inequality-based equations are solved to ensure that the bulk compound remains in thermodynamic equilibrium and no other alternative phases are formed, leading to specific allowed chemical potential value(s) of each species referenced to its standard state.} The reference phases are typically collected from the literature or databases such as the Materials Project \cite{Jain2013}, requiring several more computations using the same level of theory and convergence parameters as the defect simulations. The saving grace here is that reference calculations may already be available, though typically only at semi-local levels of theory, and thus not contribute to the overall computational expense. \\

    \item \underline{Defect complexation}: Many point defect calculations are incomplete without consideration of complexes that are likely to form and alter defect behavior. In many compounds such as Ga$_2$O$_3$, CdTe, and SiC, double and triple defect clusters have been found to be important, often providing a better interpretation of experimental results than singular defects alone \cite{Ga2O3-complex,CdTe-complex,Lee-SiC}. There are an arbitrary number of defect complexes that may form in a lattice (given the set of individual intrinsic and extrinsic defects), and the need to simulate likely candidates adds further to the cost of defect simulations. As an example, \textbf{Figure \ref{fig:Intro_Figure}d} shows a double vacancy complex along with the single vacancies in SiC \cite{Lee-SiC}. \\

    \item \underline{Vibrational properties}: Defects are typically simulated using static athermal calculations, neglecting all temperature-dependent contributions to free energy beyond configurational entropy. In many cases, however, finite-temperature effects can significantly impact defect energetics, notably in materials processed at high temperatures; including ceramics, refractory materials, and metals. Vibrational entropy is typically the dominant temperature-dependent factor beyond configurational entropy, but requires simulation of lattice dynamics in large defect-containing supercells which can be exorbitantly expensive. Employing the harmonic phonon approximation and neglecting anharmonicity and volume expansion can minimize the cost of these additional calculations while still achieving significant accuracy improvements over athermal calculations \cite{mosquera-lois_point_2025}. \textcolor{black}{Additionally, accelerated ab initio molecular dynamics could be performed to determine properties such as the thermal expansion using quasi-harmonic approximation.} \textbf{Figure \ref{fig:Intro_Figure}e} pictures the construction of force constants from a defect supercell which could be significantly accelerated using ML \cite{zhou2025onedefectpotentialstrategy}. \\
    
\end{itemize}

Each of these factors is important for the accuracy of defect simulations, but also comes with compounding computational demands which limit the pace and scope of defect modeling. There are an increasing number of examples in the literature of data-driven and AI/ML approaches being used to accelerate calculations and address these issues. With the availability of accurate surrogate models and potentials, major bottlenecks could be eliminated and the process of going from a pristine structure to the complete defect energy picture made substantially more seamless. In the next sections, we will discuss recent data-driven studies on understanding and screening of defects in solids, including key chemical and structural descriptors that have been shown to correlate with computed defect properties and yield surrogate models for quick estimation. We will also delve into how ML force fields are being used to accelerate different aspects of defect computations listed above, and finally, relate these efforts to experiments. \\

\section*{Descriptor-based empirical relationships and predictive models for defect properties}

The earliest and most routine applications of ML for materials involved regression or classification models based on unique vectorial ``descriptors". Also known as features or representations, these descriptors are easy-to-obtain physical or chemical characteristics of constituent atomic species, information about their relative structural arrangements in materials, or surrogate properties that correlate with more expensive properties \cite{Rampi-ML,Schmidt-ML}. Just as important as the quality and quantity of the training data here is the actual choice of descriptors -- namely their ability to represent the key physical behavior -- which is dictated by domain expertise understanding of specific factors that most influence a property of interest, and by the need for parsimony. The pervasiveness of oxide compounds in the computational literature means that an oversized number of ML efforts in materials have focused on oxides, with an eye to optimizing electronic, dielectric, piezoelectric, catalytic, or other properties. Existing DFT datasets of binary and ternary oxides have proven an ideal playground for testing a variety of descriptor-based ML approaches, and have also led to the creation of many defect datasets which involve oxygen vacancies, substitutional cation-site dopants, or other defects within oxide lattices. \\

Some of the most common efforts in accelerating defect predictions revolve around estimating the oxygen vacancy formation energy (E$^f$(V$_O$)), which is crucial for solid oxide fuel cells, catalytic performance for water splitting and CO$_2$ reduction, and application in scintillators and as transparent conductors. Multiple DFT datasets of E$^f$(V$_O$) computed for a variety of oxides are available in the literature, ranging from tens to thousands of data points. Interestingly, several empirical formulae and regression models have been developed for predicting E$^f$(V$_O$) for different types of oxides, with the most important features found to be the oxide formation enthalpy, band gap, O 2$p$ band center energies, and electronegativity differences \cite{ML-Deml-1,ML-Deml-2,ML-Wan,ML-Wexler,ML-Liu,ML-Wang,ML-Pentyala,kumagai_insights_2021,ML-Liu,kiyohara_machine_2025}. These studies largely focus on screening oxides for \textcolor{black}{various} applications based on a desired range of neutral V$_O$ formation energy values, with charged calculations and compensating defects often left for future work. \\

\begin{figure*}[ht]
\centering
\includegraphics[width=\linewidth]{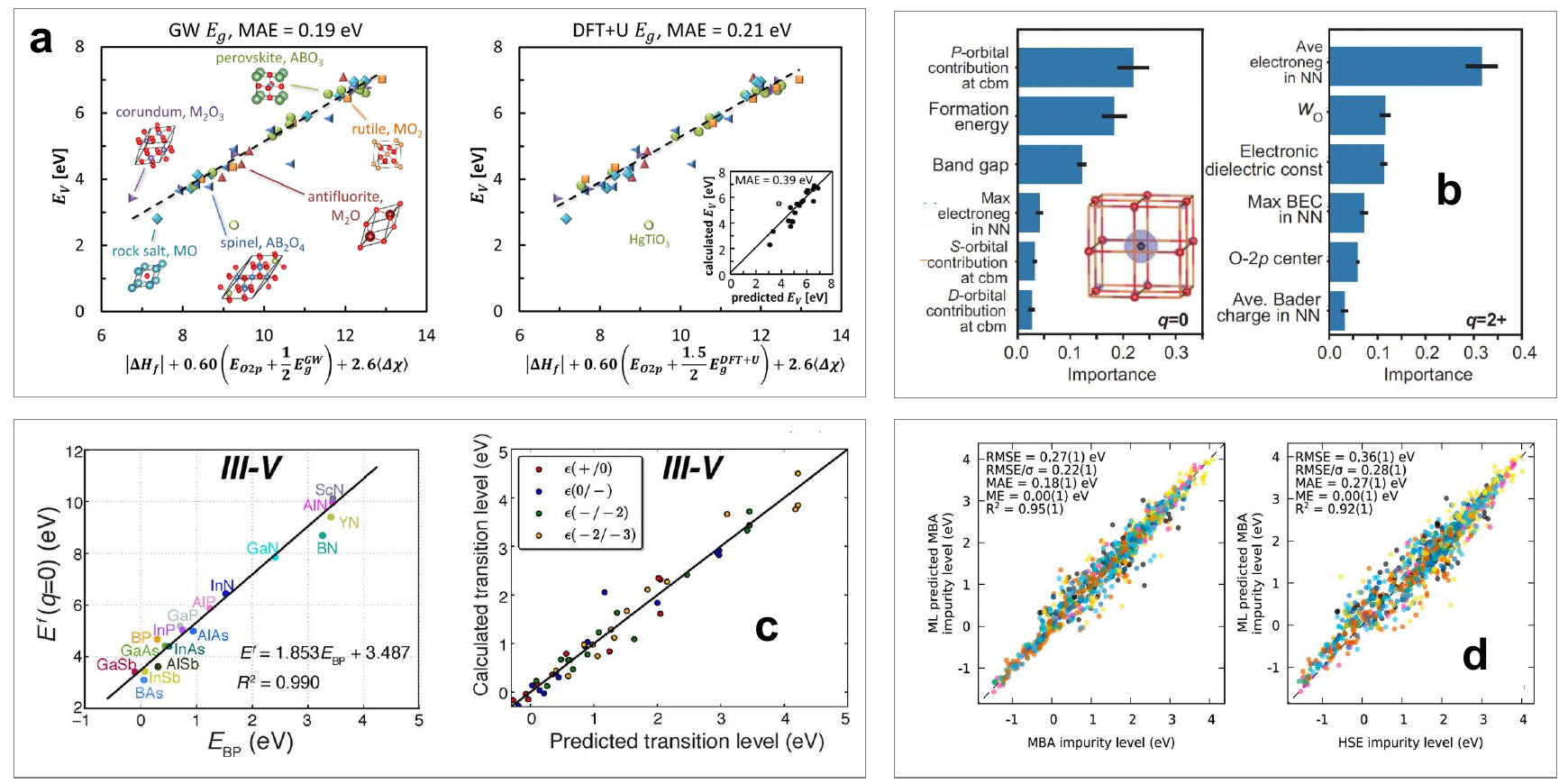}
\caption{\textbf{Descriptor-based ML models for defect predictions.} \textbf{(a)} Equations for predicting oxygen vacancy formation energy from the oxide formation enthalpy, electronegativity difference, O 2p band center, and band gap from either GW or DFT+U. As shown in the inset, predictions become worse for new out-of-sample data points. Reproduced with permission from Deml et al. \cite{ML-Deml-2}. (b) Descriptors ranked by order of importance for predicting oxygen vacancy formation energy in neutral and 2+ charge states. Reproduced with permission from Kumagai et al. \cite{kumagai_insights_2021}. (c) Predicting cation vacancy defect energies for binary III-V semiconductors: correlation between neutral defect formation energy and branch point energy, and a model for predicting the defect charge transition level. Reproduced with permission from Varley et al. \cite{ML-Varley}. (d) ML predictions of ``modified band alignment" levels compared with impurity levels from HSE06. Reproduced with permission from Polak et al. \cite{Polak-ML}.}
\label{fig:Desc_ML}
\end{figure*}

For instance, as far back as 2014, Deml et al. developed simple linear functions combining oxide formation enthalpy, band gap (from different functionals), O 2$p$ band center, and electronegativity difference between O and the cation atoms, leading to formation energy predictions with mean absolute error (MAE) of around 0.2 eV \cite{ML-Deml-1,ML-Deml-2} for neutral $V_O$. Models were trained on a small dataset of less than 50 points but quickly extended to over 2000 compounds where the descriptors were known and available, leading to candidates with desired vacancy formation. A fascinating aspect of these efforts from Deml et al. is that multiple useful functions were derived; e.g., band gaps computed at GW accuracy yielded a model with MAE of 0.19 eV, but a model using the much cheaper DFT+U band gap (the Hubbard U correction made necessary from the prevalence of transition metal cations) instead shows a nearly equally good MAE of 0.21 eV -- for \emph{neutral} $V_O$. A performance comparison of the models developed by Deml et al. using either type of band gap is presented in \textbf{Figure \ref{fig:Desc_ML}a}, along with some of the representative oxides in the training dataset. Interestingly, E$^f$(V$_O$) predictions for 18 out-of-sample oxides showed a larger MAE of 0.39 eV, as shown in the inset of the right side plot of \textbf{Figure \ref{fig:Desc_ML}a}, implying the need for further model refinement. \\

In the years since, Wan et al., Wexler et al., Kumagai et al., Liu et al. and Kiyohara et al. \cite{ML-Wan,ML-Wexler,kumagai_insights_2021,ML-Liu,kiyohara_machine_2025} have derived further data-driven models for predicting E$^f$($V_O$) in various oxides, using combinations of physical properties of constituent cations, structural features based on ionic radii, and energetic features such as crystal binding energy and energy above hull, by mining datasets of a few thousand compounds. The best MAE values for E$^f$(V$_O$) predictions reported in these works are $\sim$ 0.4 eV by Wan et al. \cite{ML-Wan}, 0.45 eV by Wexler et al. \cite{ML-Wexler}, 0.34 eV by Kumagai et al. \cite{kumagai_insights_2021}, and $\sim$ 0.3 eV by Kiyohara et al. \cite{kiyohara_machine_2025}. Kumagai et al. in 2021 even extended their work beyond neutral V$_O$ to the 2+ charge state, which is the most commonly exhibited state by O vacancies; MAE was slightly higher at 0.44 eV for 2+ E$^f$(V$_O$) \cite{kumagai_insights_2021}, and \textbf{Figure \ref{fig:Desc_ML}b} shows the specific descriptors that respectively contributed the most to predicting neutral and 2+ E$^f$(V$_O$), showing some key differences. More recently, Kiyohara et al. in 2025 also extended their work to charged vacancies and found errors to increase compared to neutral E$^f$(V$_O$) predictions for different types of models \cite{kiyohara_machine_2025}. \textcolor{black}{The range of E$^f$(V$_O$) prediction errors reported here is a factor of the different training dataset sizes, supercell sizes, and types of descriptors used in these works; moreover, charged vacancy predictions are harder, pointing to the need for additional descriptors. Additional limitations stem from low levels of theory (i.e. semi-local DFT) and limited consideration of magnetic states (in transition metal oxides).} \\

Instead of hand-crafted descriptors, Witman et al. \cite{witman_defect_2023} employed a deep graph neural network model in the same year, based on the crystal graph convolutional neural network (CGCNN), \cite{xie_crystal_2018} for the same goal of predicting E$^f$($V_O^0$) to enable high-throughput materials discovery for STHP. As evidence for model generalizability, they showed that MAE for E$^f$($V_O^0$) increases from \SI{0.3}{eV}, when predicting for \emph{unseen} vacancy sites in \emph{seen} compositions with other \emph{seen} E$^f$($V_O^0$), to \SI{0.5}{eV} for E$^f$($V_O^0$) in \emph{unseen} compositions (with \emph{seen} cations), to \SI{1.2}{eV} for completely unseen cations. Here `seen/unseen' refers to whether those cations/compositions/vacancies have been included in the training data or not (i.e. if the model has `seen' them or not). They also noted how formation energy predictions for other point defects had MAEs roughly a factor of two larger than E$^f$($V_O^0$), likely related to a greater spread in formation energies and \textcolor{black}{availability of fewer data points} per defect type. Ultimately, their conclusion was that the primary limiting factor in model accuracy was \textcolor{black}{scarcity in} the training dataset size and diversity. These authors and collaborators recently extended the work in 2025, using a similar CGCNN-based architecture trained on DFT nudged elastic band (NEB) calculations to predict vacancy migration barriers \cite{way_defect_2025}. A key finding was that the overall MAE of \SI{\sim0.9}{eV} for migration barriers is skewed by low accuracies for extreme energy barriers ($\geq$\SI{4}{eV}), with a significantly lower MAE of \SI{\sim0.5}{eV} for energy barriers between $1-$\SI{4}{eV}, which encompasses most migration paths. Recognizing where quantitative accuracies for ML models are crucial, where qualitative accuracy instead suffices, and assessing them accordingly will be important for establishing both their utility and their limits. \\

One of the earliest efforts in \textcolor{black}{making charged defect energy predictions} that we are aware of comes from Varley et al. \cite{ML-Varley} in 2017, who developed equations to predict the (neutral) E$^f$ and the transition levels $\epsilon$(q$_1$/q$_2$) of cation vacancies in II-VI and III-V semiconductors using only specific bulk properties, namely the branch-point energy, the dielectric mid-gap energy, and the interstitial hydrogen transition level. The +1/-1 transition level of a H$_i$ defect in the material can be universally aligned to the standard hydrogen electrode potential at 4.4 eV below the vacuum level, and thus has a clear correlation with the band edges and expected charge transitions of point defects. Simple models were thus derived to yield locations of $\epsilon$(0/-1), $\epsilon$(-1/-2), and other transitions of V$_M$ defects (where M is the cation), and deployed for predicting these levels in many alloy systems such as ZnSeTe and ZnMgO. \textbf{Figure \ref{fig:Desc_ML}c} shows, for III-V semiconductors, how the branch point energy correlates with the neutral cation formation energy, and the accuracy of predictive models for different charge transition levels. \\

Inspired by the work from Varley et al., in 2020 Mannodi-Kanakkithodi and Chan trained regression models on DFT datasets of the neutral formation energies and charge transition levels of hundreds of native and extrinsic defects across binary zincblende-phase group IV, III-V, and II-VI semiconductors \cite{Mannodi-npj,Mannodi-Patterns}. These models used information about the identity and well-known physical, electronic, and energetic properties of the host compound, the defect site, and the defect species, to directly predict the complete charged defect formation energy picture at cation-rich and anion-rich conditions. This work was extended to screen for low-energy dopants and impurities and predict the shallow or deep nature of their energy levels in the band gap. One of the main findings in these works, similar to the observation from Varley et al. \cite{ML-Varley}, was the utility of cheaper unit-cell properties in predicting expensive supercell defect energies. This goes to the heart of ML efforts for materials design, where the ultimate goal is to recognize easy representations and surrogate quantities that can predict something much more computationally intensive. The same concept was used by Mannodi-Kanakkithodi and Chan to train models for defect predictions in Pb-based hybrid perovskites, resulting in the discovery of several low energy Pb-site dopants with mid-gap levels which may be useful for intermediate band solar cells, as well as the formulation of simple empirical relationships to yield defect formation energies and charge transition levels \cite{Mannodi-JMS}. \\

In 2021, Polak et al. developed a ``modified band alignment" (MBA) method for estimating HSE06 (hybrid DFT) impurity levels in III-V semiconductors using the calculated levels from a lower theory (LDA or GGA functional), the shift in the electrostatic potential, the band gap, and an empirical constant obtained by fitting to known data \cite{Polak-MBA}. This was extended to an ML model which combined inputs including elemental properties such as ionization energies and electronegativities, and the impurity position and charge state, to accelerate the HSE06 prediction either via the MBA method or via direct prediction \cite{Polak-ML}. The ML-MBA level has a root mean square error (RMSE) of 0.27 eV compared to true MBA values and an RMSE of 0.36 eV compared to the HSE06 level, as pictured in \textbf{Figure \ref{fig:Desc_ML}d}. The model was applied to both III-V and II-VI semiconductors and deployed for predictions across nearly 35,000 impurity charge transition levels. Mannodi-Kanakkithodi et al. showed the \textcolor{black}{potential utility} of ``\textcolor{black}{delta} learning" for defect predictions \cite{Mannodi-npj} \textcolor{black}{with limited high-fidelity dataset sizes: including} PBE defect levels \textcolor{black}{as} descriptors dramatically improved HSE06 defect level prediction from a regression model. 
One crucial limitation worth noting with any multi-fidelity or `delta-learning' approach for defects, however, is that these models will usually not be able to rectify any \emph{qualitative} errors in the lower-level calculations (such as incorrect geometries \cite{squires_guidelines_2026} or localization \cite{kumagai_insights_2021}), but can only be expected to correct \emph{quantitative} inaccuracies (such as defect level positions, but not whether a shallow level from semi-local DFT is actually deep). 
\textcolor{black}{Multi-fidelity or delta learning can thus boost dataset size and coverage by leveraging low-fidelity data that can be more readily computed,\cite{wang_multifidelity_2026} but is expected to have limited value (if not detrimental effects) when the low-fidelity data comprises qualitative errors.}
\\

In 2023, Wu et al. \cite{ML-Wu} applied a text data mining approach to collect DFT-computed defect levels in perovskite semiconductors from the literature, and trained regression models on the dataset of 771 points to predict defect levels with an RMSE of roughly 0.3 eV. The input descriptors involved a variety of elemental, ionic, electronic, and compositional features. The best model was used to predict transition levels for all intrinsic defects in two Cs-based inorganic perovskites, Cs$_3$Sb$_2$Br$_9$ and Cs$_2$SnBr$_6$, and the predictions were validated with GGA calculations. In 2025, Khamdang et al. \cite{ML-Khamdang} generated a smaller dataset of Sn-site and Cs-site dopants in orthorhombic CsSnI$_3$ using high-accuracy HSE06 computations with spin-orbit coupling (SOC), and trained regression models to predict charged formation energies utilizing a variety of elemental features and bulk structural and thermodynamic properties. A combination of ML prediction and targeted DFT was used to determine suitable dopants such as Y, Sc, La, and Zr, which may suppress p-type doping in CsSnI$_3$. 
These studies can be taken as proofs of concept for the utility of data-driven approaches in defect modeling for rapid estimation of formation energies for discovery or screening purposes. \\

\section*{Defect predictions with machine learning force fields}

Machine-learned force fields (MLFFs), also known as machine-learned interatomic potentials (MLIPs), are becoming an indispensable tool for computational materials science \cite{behler_generalized_2007,mroz_crossdisciplinary_2025,kulichenko_data_2024}. A wide variety of architectures and approaches have emerged in this rapidly-expanding field, with several review articles and practical guides in the literature \cite{deringer_machine_2019,morrow_how_2023,jacobs_practical_2025,unke_machine_2021}. These are ML models trained on datasets of quantum-mechanical simulations, with inputs of atomic structure (geometry and composition), and outputs of total energies and atomic forces -- in some cases extended to include charge, spin or other properties. 
As such, they aim to directly replace DFT or other electronic structure methods as the energy calculator for at least some portion of the computational workflow, thereby reducing computational costs and allowing far larger system sizes, longer timescales, or wider configurational space. This gets at a clear advantage of MLFFs over descriptor-based approaches; they are much more flexible.
Instead of being trained to predict single quantities such as formation energies, they can be used to predict all properties which can be derived from total energies and atomic forces (and potentially other outputs), which includes formation energies, ground-state and metastable geometries, phonons and molecular dynamics, binding energies, migration pathways, and more. Of course, this generality only becomes useful when the models themselves are sufficiently accurate. \\

The defects, materials informatics, and MLFF development research communities recognize defects as a key application area of MLFFs \cite{cusentino_explicit_2020,deng_systematic_2025,jacobs_practical_2025,shuang_universal_2025,wines_chipsff_2025,rinaldi_noncollinear_2024}.
From a defect modeler's perspective, MLFFs offer to reduce the large computational costs which restrict the scope and pace of defect research, discussed above. From the perspective of the MLFFs community, this computational cost, along with established modeling workflows and importance to material properties, makes defects a natural extension beyond bulk crystal modeling where MLFFs can be usefully applied – a category which surfaces, interfaces, ionic migration, and large-scale molecular dynamics also fall under. In many cases thus far, however, defects have been used as a perfunctory example application of MLFFs trained on (mostly) bulk crystal data, without much thought given to the accuracy or utility of the predictions. For instance, an unfortunately common occurrence is the reporting of mean energy-\emph{per-atom} error metrics on defect datasets -- as is standard for bulk systems -- and concluding that good accuracy has been achieved when these values are comparable to those for pristine crystals. However, such analysis overlooks the fact that most of the defective simulation cell is still `pristine', giving a size-dependent metric where large errors in defect formation energy can be washed out when dividing by the total number of atoms \cite{cusentino_explicit_2020}. We stress the need for prudence and pragmatism in the application of MLFFs to defects – and of course other application areas beyond bulk crystals. Rigorous examination and benchmarking of accuracies, alongside clear recognition of their limits, will be crucial for building confidence in these approaches within this community. \textcolor{black}{We note that a large number of MLFF architectures have been proposed in recent years, with developments still ongoing, and so the specific models discussed here represent only a fraction of available MLFFs.\cite{riebesell_framework_2025} Future developments may well lead to alternative ML architectures achieving top performance for MLFFs, for defect simulations and beyond.} \\

\begin{figure*}[ht]
\centering
\includegraphics[width=\linewidth]{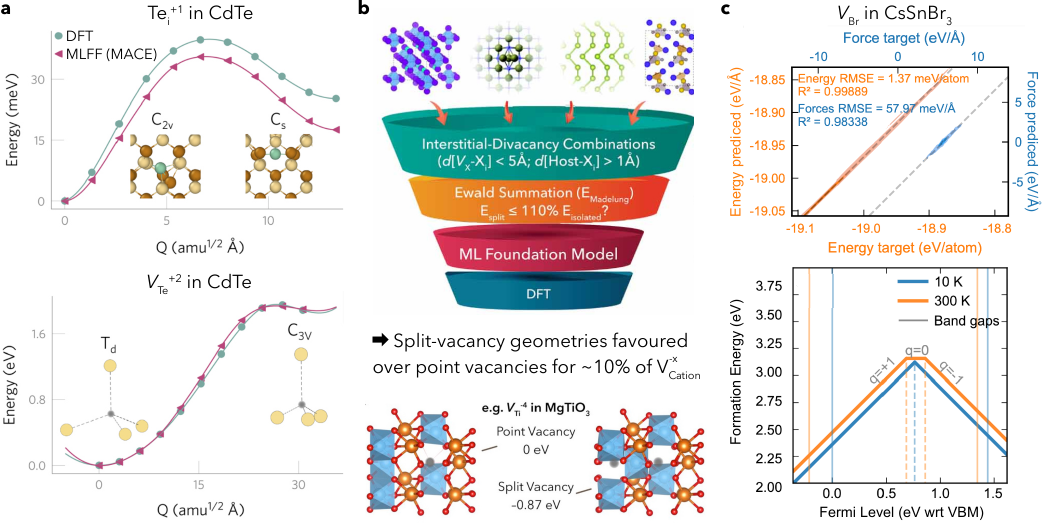}
\caption{\textbf{Applications of machine-learned force fields (MLFFs) in defect modeling.} \textbf{(a)} Potential energy surfaces of defect configurations identified with structure searching \cite{mosquera-lois_shakenbreak_2022} for Te$_{\textrm{i}}^{+1}$ (top) and $V_{\textrm{Te}}^{+2}$ in CdTe, calculated using DFT and a trained MACE MLFF \cite{batatia_mace_2022}. The $x$-axis variable $Q$ represents a mass-weighted displacement coordinate that tracks the structural change along the pathway between defect configurations. Adapted with permission from Mosquera-Lois et al. \cite{mosquera-lois_point_2025}
\textbf{(b)} ML-accelerated screening workflow to predict the formation of split cation vacancies for all $\sim$150,000 compounds in the Materials Project database \cite{jain_materials_2013}. Underneath is an example of relaxed point and split vacancy geometries for $V_{\textrm{Ti}}^{-4}$ in \ce{MgTiO3}, with Ti (polyhedra) in blue, Mg in orange, O in red, and vacancies as semi-transparent grey circles. Adapted with permission from Kavanagh \cite{kavanagh_identifying_2025}.
\textbf{(c)} Modeling finite-temperature effects on formation energies and transition levels for $V_{\textrm{Br}}$ in \ce{CsSnBr3}. The upper parity plot shows the model accuracy for energies and forces, while the lower panel shows the temperature-dependent defect formation energy diagram with the Fermi level referenced to the valence band maximum. Dashed and solid vertical lines indicate the charge transition levels and band edge positions, respectively, at 10 (blue) and \SI{300}{K} (orange).
Adapted with permission from Hainer et al. \cite{hainer_thermal_2025}.}
\label{fig:MLFFs_Figure}
\end{figure*}

One natural application area of MLFFs within the defect workflow is for potential energy surface (PES) exploration, illustrated in \textbf{Figure \ref{fig:MLFFs_Figure}(a,b)}, as is becoming common within catalysis and adsorption simulations \cite{jung_machinelearning_2023,lim_accelerating_2025}.
Defects can adopt multiple locally-stable geometries with vastly different formation energies and behavior \cite{mosquera-lois_identifying_2023}, necessitating global optimization strategies to ensure the correct ground-state and metastable structures are identified.
Within this structure-searching step, quantitative accuracy becomes less important, where only the qualitative prediction of locally-stable structures and, ideally, their relative energy ordering are required.
In 2021, Arrigoni and Madsen \cite{arrigoni_evolutionary_2021} showed how a Gaussian Process (GP) surrogate model could be employed to partially replace DFT and reduce computational cost within an evolutionary algorithm approach for refining candidate defect geometries. They applied this model to Silicon self-interstitials and $V_O$ in anatase-\ce{TiO2}. GP models are typically more limited in accuracy than deep neural networks, but a major benefit is the built-in uncertainty metric provided by the posterior covariance matrix \cite{deringer_gaussian_2021}. This allows them to be seamlessly employed in active-learning workflows; where the GP model is used to evaluate candidate structures for which uncertainty is low, while DFT is called only for high-uncertainty configurations. These DFT calculations are then added to the training data and the model is re-trained `on-the-fly'. GP models have been used in the simulation of amorphous materials for some time \cite{deringer_origins_2021}, where the combination of large computational cost with many different but closely-related local atomic environments has motivated their use. Recent work from Morrow et al. in 2024 \cite{morrow_understanding_2024} extended these approaches with teacher-student model distillation to yield ultra-fast MLIPs, capable of modeling large million-atom cells and reliably quantifying dangling- and floating-bond `defects' in amorphous Silicon. \\

While Gaussian Process models have their advantages, message-passing graph neural network (GNN) models have by and large emerged as the current MLFF architecture of choice. State-of-the-art GNN models typically employ invariant or equivariant features to reflect physical symmetries \cite{batzner_advancing_2023,batatia_design_2025}, and have achieved the highest accuracies for large materials and molecular datasets thus far \cite{riebesell_framework_2025,levine_open_2025}. In 2024, Mosquera-Lois et al. \cite{mosquera-lois_machinelearning_2024} took the pre-trained M3GNet \cite{chen_universal_2022} invariant GNN model and fine-tuned it on hybrid DFT relaxations of cation vacancies in metal chalcogenides, generated using the \texttt{doped} \cite{kavanagh_doped_2024} and \texttt{ShakeNBreak} \cite{mosquera-lois_shakenbreak_2022} defect toolkits. The authors found that even with this small and relatively biased fine-tuning dataset, the MLFF was able to successfully identify energy-lowering structural reconstructions for unseen defects in unseen compositions for \SI{90}{\%} of cases.
As a proof of concept, the authors also showed that a model trained on end-member compositions (CdTe and CdSe) could identify stable geometries in a disordered mixed system (CdSe$_x$Te$_{1-x}$) with diverse local coordination environments. \\

In recent work from Kavanagh \cite{kavanagh_identifying_2025}, it was found that leading `foundation potentials' \textcolor{black}{(which are highly-transferrable MLFFs pre-trained on large quantities of diverse atomistic data)} were capable of identifying stable split-vacancy geometries across diverse material classes (\textbf{Figure \ref{fig:MLFFs_Figure}b}), even without fine-tuning. 
\textcolor{black}{In this case, the foundation potentials used were based} on the MACE \cite{batatia_mace_2023} and NequIP \cite{tan_highperformance_2025} equivariant GNN architectures\textcolor{black}{, trained on large datasets of bulk material simulations}.
Crucially, this surprising performance was enabled by the restriction to fully-ionized charge states \cite{squires_guidelines_2026} where split vacancies have mostly been observed \cite{fowler_metastable_2024}, having simpler energy surfaces dominated by electrostatics and strain with minimal charge localization. A similar machine-learning assisted search for defect complexes was performed by Jiang et al. in 2024 for \ce{ThO2} \cite{jiang_machine_2024}, discovering unexpectedly high levels of metastability and polymorphism, along with counter-intuitive non-compact configurations. More generally, in the past five years MLFFs have been used to model defects in metal chalcogenides \cite{rahman_defectff_2025,rahman_understanding_2024,wang_multifidelity_2026,mosquera-lois_point_2025}, halide vacancies and interstitials in \ce{Cs3PbI3} and \ce{Cs3PbBr3} \cite{pols_how_2023}, dislocations and vacancies in metals \cite{goryaeva_efficient_2021,wen_specialising_2021}, oxygen vacancies in amorphous silica \cite{li_machine_2026}, and to predict vacancy formation energies across nearly $\sim100,000$ compounds \cite{berger_screening_2025a}. 
\textcolor{black}{These approaches have been successful in extending predictive capabilities within these studies, allowing more exhaustive defect energy surface exploration, more comprehensive defect enumeration in amorphous/disordered systems and far higher throughput in screening formation energies across materials space, than would be possible with conventional quantum-mechanical methods. However, these defect MLFF studies have thus far been either restricted to small sets of defects in single compositions, or to high-throughput screening but with low ‘semi-quantitative’ accuracy. It is hoped that future work in this area will allow the application of MLFFs to diverse defect/material spaces while retaining high accuracy, as has been achieved in related materials sub-fields.\cite{riebesell_framework_2025,pota_thermal_2024,tran_open_2023,moon_catbench_2025}}
\\

Another frontier in defect modeling, which MLFFs promise to unlock, is the consideration of finite-temperature effects on free energies, charge transition levels, and other properties such as migration and carrier recombination \cite{cheng_computing_2018}. Direct simulation of dynamic defect properties has mostly remained beyond the reach of first-principles methods, due to exorbitant costs, with defect modeling mostly limited to static calculations. In 2025, Mosquera-Lois et al. \cite{mosquera-lois_point_2025} showed how MLFFs could be used to explore such dynamic defect behavior and directly quantify finite-temperature contributions to the free energy of formation.
Using MACE \cite{batatia_mace_2022} MLFFs to efficiently reproduce the DFT defect energy surfaces (\textbf{Figure \ref{fig:MLFFs_Figure}a}), the authors showed that low-energy metastable configurations of Te$_i^{+1}$ in CdTe \cite{kavanagh_impact_2022,kavanagh_accurately_2024} were highly populated at room temperature, and that thermal effects -- dominated by vibrational, spin and configurational entropy contributions -- increased the predicted concentrations by two orders of magnitude.
The effects on $V_{Te}^{+2}$ were less significant; due to its harder PES, static behavior, and lack of low-energy metastable states; showing the defect-dependent impact of finite-temperature effects. The authors employed a similar approach to examine $V_{Cl}$ in \ce{CsPbCl3} under different charge states \cite{mosquera-lois_dynamic_2025}, finding that despite significant dynamical effects and oscillations in optical transition levels, non-radiative recombination and thermodynamic transition levels remained consistent with the static defect formalism. \\

In another work appearing in late 2025, Hainer et al. \cite{hainer_thermal_2025} used neuro-evolution potential (NEP) MLFFs to model finite-temperature effects on vacancy formation energies and charge transition levels in MgO, LiF, and \ce{CsSnBr3} (\textbf{Figure \ref{fig:MLFFs_Figure}c}), notably predicting the neutral charge state of $V_{\textrm{Br}}$ to be stabilized at $T > \SI{60}{K}$.
Very recently, Zhu et al. \cite{zhu_predicting_2025} investigated thermal effects on native defects in GaAs, fine-tuning the MACE-MP0 foundation potential and training an equivariant GNN for the DFT Hamiltonian (DeepH-E3 \cite{gong_general_2023}) to predict electronic structure. The authors showed how an active-learning approach could be employed to rapidly construct and validate these models, to minimise computational costs. Future work in this area, enabled by MLFFs, will yield more general insights for the community on the impact of finite-temperature effects on defect properties; such as when they are likely significant, and when they can be safely neglected. \\

The modification and deployment of MLFFs for defect studies is a burgeoning area of research, with a number of applications in recent years. For instance, Dramko et al. \cite{dramko_ADAPT_2025} recently proposed a transformer architecture which replaces GNN graph representations with a direct coordinates-in-space formulation, to avoid issues with over-smoothing and long-range interactions \cite{yan_case_2024}. This resulted in significant improvements to prediction accuracies compared to state-of-the-art GNN MLFFs. To similarly avoid over-smoothing and explicitly bias the model toward defects, Yang et al. introduced a defect-informed GNN architecture \cite{yang_modeling_2025}, but to predict relaxed defect structures rather than energies and forces as with MLFFs. Both of these studies appeared in 2025, and we anticipate many developments in methodologies for defect MLFFs over the next decade. \\

Looking forward, we note that the emergence of defect-ready MLFFs calls for the development of advanced methodologies and software capable of leveraging these accelerations and streamlining the dataset generation, training, and deployment workflows. 
\textcolor{black}{Specifically, defect MLFFs will require simulation toolkits which can efficiently automate thousands of calculations (such as \texttt{atomate2}\cite{ganose_atomate2_2025} or \texttt{AiIDA}\cite{huber_aiida_2020}) and parse their results (such as \texttt{doped}\cite{kavanagh_doped_2024}, \texttt{pydefect},\cite{kumagai_insights_2021} or \texttt{DASP}\cite{huang_dasp_2022}), to fully benefit from their accelerated calculation throughput. Extensions of these automated toolkits to advanced defect properties, such as non-radiative recombination, luminescence lineshapes and finite-temperature effects, is an active area of development which will be key to achieving the advanced predictive capabilities promised by ML-accelerated defect simulations.\cite{hainer_thermal_2025,mosquera-lois_dynamic_2025} We stress that interoperability in these computational toolkits, as offered by the Python computational materials science ecosystem, is crucial for minimising redundancy, allowing practitioners to readily combine various advanced approaches with minimal human effort. Moreover, the continued development of toolkits which implement established methodologies for MLFF data generation, training and validation -- such as \texttt{MLIP arena}\cite{chiang_mlip_2025} or \texttt{autoplex}\cite{liu_automated_2025} -- will aid the ready adoption of MLFFs within the defect community by reducing the human effort and ML expertise requirements for users. Alongside,} the provision of open-access defect datasets \cite{huang_unveiling_2023,angsten_elemental_2014,björk_twodimensional_2024} will allow more rapid evaluation of alternative MLFF architectures, in efforts to resolving remaining issues with defect MLFFs. 
\textcolor{black}{Another important factor for the successful adoption of defect MLFFs will be the development of robust and physically-relevant prediction benchmarks, which has proven valuable (and non-trivial) in other sub-domains of computational materials science.\cite{riebesell_framework_2025,pota_thermal_2024,tran_open_2023} These will allow the realistic assessment of MLFF predictive accuracies for defect properties and to help guide future advances in this research space.}
\\

\section*{ML prediction of defect phonons}

\begin{figure*}[ht]
\centering
\includegraphics[width=\linewidth]{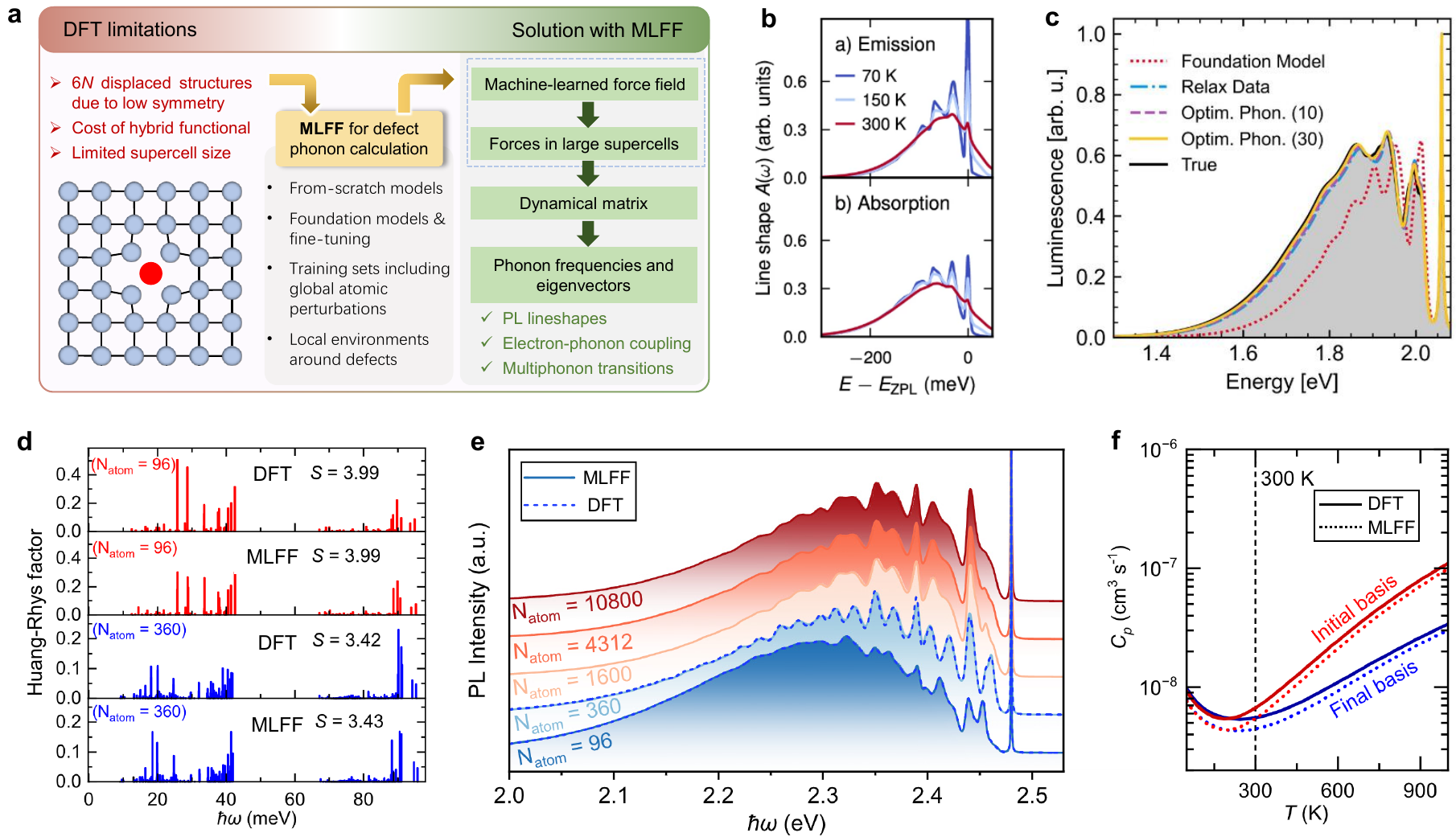}
\caption{\textbf{Applications of machine-learned force fields (MLFFs) in defect phonon calculations.} \textbf{(a)} Advantages of MLFFs for phonon calculations in defect-containing supercells. \textbf{(b)} Photoluminescence (PL) lineshapes from an ML potential using the ACF approach. Reproduced with permission from Linderälv et al. \cite{linderalv2025optical}. \textbf{(c)} PL lineshapes obtained from foundation models and from models after fine-tuning. Reproduced with permission from Turiansky et al. \cite{turiansky2025machinelearningphononspectra}. \textbf{(d)} A comparison of Huang-Rhys factors calculated from DFT and from MLFFs. \textbf{(e)} PL lineshapes from MLFF models trained from scratch. \textbf{(f)} Nonradiative capture coefficients from MLFF modelss trained from scratch. Plots in d, e, and f are reproduced with permission from Zhou et al. \cite{zhou2025onedefectpotentialstrategy}.}
\label{fig:MLFFs_Phonons}
\end{figure*}

The previous section reviewed the progress of MLFFs in predicting defect formation energies and local structural relaxation. These developments have demonstrated that MLFFs are highly effective for modeling the static properties of defects. However, many central defect-related processes critically depend on defect phonons, including thermal conductivity, electron-phonon coupling, and phonon-assisted electronic transitions. Accurate modeling of defect-induced vibrational properties therefore represents a natural and necessary extension of MLFF applications. Compared with perfect crystals, vibrational properties in defective systems involve strong local lattice reconstructions, long-range elastic distortions, and charge-state–dependent structural responses. As a result, traditional DFT calculations of phonons in large defect-containing supercells are computationally prohibitive. To overcome this barrier, several recent studies have explored the use of MLFFs as efficient surrogates for computing force constants, vibrational modes, and phonon scattering processes in defect-containing supercells. The advantages of using MLFFs for defect phonon calculations compared to DFT are presented in \textbf{Figure \ref{fig:MLFFs_Phonons}a}. \\

An important early example is the work of Shimizu et al. in 2022 \cite{Shimizu2022}, who constructed a Behler–Parrinello neural network potential (NNP) for the nitrogen vacancy V$_\text{N}$ in GaN. A key modification is the inclusion of an additional system-charge node in the input layer, which allows a single NNP to consistently describe phonon properties of defects across different charge states (0, +1, +2, +3). \textcolor{black}{Their model reproduces the DFT phonon spectra with high accuracy, supported by low errors in energies (1.4 meV/atom) and forces (64 meV/Å), and shows close agreement in both acoustic and optical branches, with only minor deviations at high frequencies.} In a related study, Dou et al. in 2025 \cite{Dou2025} developed a Behler–Parrinello NNP for AlN to investigate defect-induced phonon scattering and thermal transport. Their model was trained on a diverse dataset that included pristine AlN and structures containing various point defects in different charge states, enabling accurate prediction of second- and third-order force constants. This allowed them to compute defect-limited thermal conductivity with DFT-level accuracy but at a fraction of the computational cost. Unlike Shimizu's approach \cite{Shimizu2022}, which explicitly introduced a charge descriptor to unify multiple charge states within a single model, Dou et al. \cite{Dou2025} relied on a conventional model trained directly on mixed neutral and charged configurations, without an explicit charge input. A related study in 2024 on cubic boron nitride (cBN) with vacancy defects and isotopic disorder also employed a conventional NNP \cite{Zhang2024}. The model was trained on pristine, vacancy-containing, and isotope-disordered configurations, enabling accurate prediction of phonon dispersions and densities of states in the presence of both point defects and mass disorder. Since this work does not have any explicit charge descriptor, it is conceptually similar to Dou's approach but applied to a different defect landscape. \\

While some of the early applications of MLFFs successfully addressed defect phonons and phonon transport, defect physics associated with multi-phonon processes, including Huang-Rhys factors, defect-induced photoluminescence spectra \cite{Alkauskas2014PL}, and nonradiative multiphonon transitions \cite{Alkaus2014nonrad, Zhou2025}, require both accurate phonon predictions and the ability to treat very large supercells. The computational demand of traditional DFT-based phonon workflows makes such calculations challenging, and MLFFs developed more recently have emerged as powerful tools in this regime. A notable attempt to accelerate defect optical spectra using MLFFs was made by Linderälv et al. in 2025 \cite{linderalv2025optical}, who combined a machine-learned potential with a classical autocorrelation function (ACF) approach. In their work, a neuroevolution potential was trained for the ground and excited electronic states of a given defect. Fourier transforming the resulting ACF yields the optical lineshape without requiring explicit normal mode analysis; a glimpse of these results is presented in \textbf{Figure \ref{fig:MLFFs_Phonons}b}. This study for the first time demonstrated that ML-accelerated PES sampling can reproduce key vibrational contributions to defect luminescence. \\

A different direction was taken by Sharma et al. also in 2025 \cite{sharma2025accelerating}, who examined how well off-the-shelf foundation models can reproduce defect phonons and optical lineshapes without any defect-specific training. Their benchmark covered seven state-of-the-art foundation models, including several E(3)-equivariant GNN force fields as well as more general deep-graph potentials. Using these universal models to compute the full dynamical matrix for 791 defects in multiple materials, they showed that modern large-scale foundation models can already capture key trends in Huang–Rhys factors and photoluminescence lineshapes with no additional training, highlighting the surprising strength of foundation model for defect vibrational physics. \\

For a more accurate calculation, universal models remain imperfect for strongly localized modes and defects involving large structural reconstructions. To address this gap, Turiansky et al. in 2025 \cite{turiansky2025machinelearningphononspectra} adopted an intermediate strategy: starting from the same foundation model, they fine-tuned the model using only a small number (tens) of DFT structures obtained along the relaxation pathway of each defect and charge state. This compact fine-tuning dataset was sufficient to dramatically improve the predicted force constants, phonon frequencies, and vibrational eigenvectors. As a result, the refined models achieved near–hybrid-DFT accuracy for Huang–Rhys factors, spectral widths, and temperature-dependent photoluminescence lineshapes. Additional improvements could be obtained by augmenting the training set with 10-30 finite-displacement structures. A comparison between photoluminescence lineshapes from foundation models and from new fine-tuned models is presented in \textbf{Figure \ref{fig:MLFFs_Phonons}c}. \\

At the same time, to address the imperfection of universal potentials, Zhou et al. \cite{zhou2025onedefectpotentialstrategy} recently introduced the “one defect, one potential” strategy to learn the local potential energy surface of a given defect. In this approach, a compact equivariant model was trained from scratch using only a small set of randomly displaced configurations centered around the defect, allowing the model to accurately reproduce phonon dispersions and Huang–Rhys factors. With the improved force-constant embedding scheme, the model is scalable to supercells exceeding 10$^4$ atoms for PL spectra calculations. Importantly, a practical machine learning method to calculate full-dimensional nonradiative multiphonon capture rate was provided. By explicitly retaining all phonon modes from MLFFs while avoiding equal-mode or reduced-dimension approximations in the transition model, this method enables quantitatively accurate evaluation of defect-assisted recombination processes at a computational cost orders of magnitude lower than brute-force DFT. Prediction comparisons for Huang-Rhys factors, PL lineshapes, and nonradiative capture coefficients are shown in \textbf{Figure \ref{fig:MLFFs_Phonons}(d-f)}. \\

\textcolor{black}{From a broader perspective, these three approaches reflect a trade-off between transferability and accuracy in modeling defect phonons. Foundation models offer high efficiency and broad applicability, but may lack the fidelity required to capture large lattice relaxations and strong electron-phonon interactions. Fine-tuning provides a practical compromise, enabling systematic improvement with limited defect-specific training data while retaining some transferability. In contrast, defect-specific models achieve the highest accuracy for a given system by explicitly learning the local potential energy surface, but at the cost of generality.
Given the intrinsic complexity of point defects, it is unlikely that a single approach will be universally optimal. Instead, the choice of strategy should be guided by the target application. For high-throughput screening or trend identification, foundation models or lightly fine-tuned models may be sufficient. For quantitatively accurate predictions of defect-related properties, such as phonon sidebands, Huang-Rhys factors, or nonradiative capture rates, defect-specific training is often necessary. Establishing such guidelines is essential to avoid the proliferation of defect-specific models without clear applicability, and to ensure that MLFF for defects form a coherent framework rather than a collection of disconnected case studies.
}\\

\section*{Connection with experiments}

When it comes to validation, the machine learning approaches discussed here are mostly benchmarked against DFT data, since training data is typically from DFT simulations, making it unreasonable to compare ML predictions directly with experiments. The validity of the ML methods is thus given by the product of their accuracy -- how well they reproduce the physical system to which their training data corresponds (e.g. the DFT energy surface) -- and the validity of that (model) system.
The physical validity and connection of defect simulations from first principles with experiments has been well studied and is touched on in other works within this collection \cite{freysoldt_firstprinciples_2014}.
Primary sources of error in contemporary defect simulations, assuming the use of accurate electronic structure methods \cite{squires_guidelines_2026,deak_accurate_2011}, tend to be remnant finite-size effects and neglect of finite-temperature contributions, with a typical expected accuracy of 0.05 - \SI{0.1}{eV} in formation energies and charge transition levels \cite{freysoldt_firstprinciples_2014,chen_accuracy_2017}.
We note that many other ML applications in science do employ experimental training data, in which case the validity of the models is dictated by the product of their accuracy and the precision of the underlying experimental measurements. However, direct measurement of defect properties are exceedingly difficult and often only possible in specific circumstances (e.g. spin-active centers) \cite{squires_guidelines_2026,jin_photoluminescence_2021,alkauskas_firstprinciples_2012,yan_distinguishing_2023,razinkovas_photoionization_2021,shepidchenko_small_2015,kavanagh_rapid_2021}, preventing the production of sufficiently large datasets for ML thus far. \\

ML-accelerated first principles simulations of defects in solids yield numerous observable properties, including ground-state and metastable defect configurations, electronic energy levels, achievable carrier concentrations and equilibrium Fermi level, as well as vibrational and other finite-temperature properties -- each of which could be, at least qualitatively, compared with measurements from suitable experiments for interpretation, understanding, and discovery. X-ray and electron diffraction patterns of structures with high concentrations of defect-related coordination environments and disorder \cite{xrd-1,xrd-2,elec-diff} can be resolved using computationally optimized defect configurations \cite{yang_modeling_2025a}, where MLFFs can help account for symmetry-breaking and thoroughly exploring the PES. 
In cases where high-resolution images of atomic structure are available, such as with Scanning Transmission Electron Microscopy (STEM) or Scanning Tunneling Microscopy (STM) analysis of two-dimensional materials, defect geometries predicted by DFT and MLFFs have been directly compared to experiment \cite{yang_modeling_2025a}.
Rojsatien et al. showed that x-ray absorption near edge structure (XANES) spectra simulated for DFT-computed low-energy defect structures can be mapped to experimentally measured spectra via linear combination fitting, to resolve the specific phases or local environments that are likely present in real-life samples \cite{xanes-1,xanes-2}. Energies or peaks measured from x-ray absorption, photoluminescence, or deep level transient or optical absorption spectra are also directly comparable with computed defect levels \cite{xas,dlts}. It is important that all possible individual defects and complexes related to native or extrinsic species have been computationally accounted for at an advanced level of theory to ensure accurate interpretation of measured levels. \\

ML-accelerated predictions of defect structures and electronic levels can be seamlessly mapped to experiments by creating a framework for ready collection of data and interpretation of results using suitable algorithms. While targeted experimental efforts for specific defects, dopants, and impurities would enable convenient comparisons, efforts must also be undertaken to efficiently acquire data reported in publications over the last several decades. E.g., Mannodi-Kanakkithodi et al. \cite{Mannodi-Patterns} manually collected measured defect levels for well-known binary semiconductors from across publications and handbooks, revealing rough general errors in DFT-computed levels in comparison. Two major challenges here are navigating the massive literature space to find relevant and reliable experimental defect results, and determining the confidence in assignments of measured level(s) to specific defects -- something most experimental techniques cannot unambiguously resolve. State-of-the-art large language models (LLMs) are very promising for automating the extraction of experimental data on defects and mapping them to predictions from DFT, surrogate models, or MLFFs, and indeed, directing DFT+ML efforts to simulate exactly what's reported in the experimental literature. \\

\section*{Conclusions}

In this perspective, we presented a short review of recent and ongoing efforts to accelerate defect simulations in solids using machine learning approaches. A variety of glittering examples from the literature show that data-driven screening, useful descriptors, and surrogate models have been instrumental in understanding the behavior of vacancies, interstitials, dopants and impurities, and in designing materials with tailored defect properties. State-of-the-art machine learning force fields enable rapid optimization of defect configurations and estimation of related vibrational and phonon properties. Many models and workflows are now published and available for use by the defect simulation community. \\

Going forward, it is vital to generate more high-fidelity simulation data for training and devise more specialized representations for structures containing defects. The majority of defect models thus far have been trained on semi-local DFT functional data, due to its computational affordability and pervasiveness in materials databases.
The failures of semi-local DFT for defect modeling are well-established, however \cite{squires_guidelines_2026,freysoldt_firstprinciples_2014,Lyons-GaN}, and so there is an increasing push to move towards hybrid functional training data for next-generation defect force fields \cite{mosquera-lois_dynamic_2025}. 
Current models can be suitably applied within active learning workflows to help generate data rationally and improve prediction accuracy. Active learning requires uncertainty estimates, which can be provided by Gaussian Process-based models or standard deviations in model ensemble prediction for neural network potentials.  We foresee this field continuing to grow rapidly in the next few years, as the use of combined DFT-ML defect simulations becomes an increasingly integral part of pipelines for rational design of electronic and energy materials. \\

\section*{ACKNOWLEDGMENTS}

A.M.K. acknowledges support from the Purdue University School of Materials Engineering faculty start-up grant, the Defense Advanced Research Projects Agency (DARPA) Young Faculty Award 2025, and the U.S. Department of Energy’s Office of Energy Efficiency and Renewable Energy (EERE) under the Solar Energy Technology Office (SETO) Award Number DE-0009332. A.M.K. also acknowledges funding provided by the Alliance for Sustainable Energy, LLC, Managing and Operating Contractor for the National Renewable Energy Laboratory for the U.S. DOE, and was supported in part by EERE under SETO Award Number 37989. M.H. was supported by the National Natural Science Foundation of China (12404089). \\

\section*{AUTHOR CONTRIBUTIONS}

A.M.K. and S.R.K. took the lead on writing and editing the manuscript. All authors contributed to the literature review and writing for different sections. \\

\section*{CONFLICT OF INTEREST}

There are no conflicts of interest to report. \\

\section*{DATA AVAILABILITY}

No new datasets were generated or analyzed during this study. \\

\bibliography{references-revised,SK_Zotero-revised}

\bibliographystyle{ieeetr}

\end{document}